\documentclass[aps,11pt,notitlepage,nofootinbib,showkeys,showpacs,prd]{revtex4}
\usepackage[utf8x]{inputenc}
\usepackage{graphicx}
\usepackage{dsfont}
\usepackage{hyperref}
\usepackage{bm}
\usepackage{dcolumn}
\usepackage[mathscr]{eucal}
\usepackage{color}
\usepackage{pst-plot}
\usepackage{amsmath}
\usepackage{comment}

\DeclareMathOperator{\iu}{\mathrm{i}}

\DeclareMathOperator{\id}{\mathds{1}}

\def\be{\begin{equation}}
\def\ee{\end{equation}}

\newcommand{\bra}[1]{\langle #1|}
\newcommand{\ket}[1]{|#1\rangle}

\newcommand{\proj}[1]{\vert #1\rangle\!\langle#1 \vert}

\begin{document}

\title{\Large \bf Disordered locality and Lorentz dispersion relations:\\ an explicit model of quantum foam
}

\author{{ Francesco Caravelli}}\email{fcaravelli@perimeterinstitute.ca}

\author{{Fotini Markopoulou}}\email{fmarkopoulou@perimeterinstitute.ca}
\affiliation{
Perimeter Institute for Theoretical Physics, \\
Waterloo, Ontario N2L 2Y5
Canada, \\and\\
  University of Waterloo, Waterloo, Ontario N2L 3G1, Canada,\\
 and\\
Max Planck Institute for Gravitational Physics, Albert Einstein Institute,\\
Am M\"uhlenberg 1, D-14476 Golm, Germany}

\medskip



\begin{abstract}
 Using the framework of Quantum Graphity, we construct an explicit model of a quantum foam, 
a quantum spacetime with spatial non-local links. The states depend on two parameters: the minimal size of 
the link and their density with respect to this length. Macroscopic Lorentz invariance requires that the quantum
superposition of spacetimes is suppressed by the length of these non-local links. We parametrize this suppression by the distribution of non-local links 
lengths in the quantum foam. We discuss the general case and then analyze two specific natural distributions. Corrections to the Lorentz 
dispersion relations are calculated using techniques developed in previous work. 
\end{abstract}

\maketitle
\newpage

\section{Introduction}
A fascinating idea proposed by Wheeler  in the early years of Quantum Gravity, is that, at the Planck scale,
geometry may be bumpy due to quantum fluctuations. This is the {\em quantum foam} \cite{WheelerFord}. While intuitively natural, this idea is very complicated to put into action.
In the present paper, we will use the framework of Quantum Graphity \cite{graphity1,graphity2,graphity3} to construct
a simple model of quantum foam. 

A key feature of a quantum foam is its non-local nature.  While non-locality is undesirable in quantum field theory, the situation in quantum gravity is open.
It is often said that the only way to cure the divergences appearing 
perturbatively in quantizations of gravity without introducing new physics (i.e., string theory or super-symmetric extensions of gravity), is to introduce some kind of non-locality in the action which smears out  Green functions evaluated on 
one point only. Until now,  ghosts in the theory have blocked  research in this direction (some progress has been 
achieved recently in \cite{stleomaz}).
For the purposes of the present work, it is important to note that there are two possible types of non-locality which contribute in different ways. One, 
violation of {\em microlocality}, disappears when  the cut-off is taken to zero, while the other,  violation of  {\em macrolocality}, or \textit{disordered locality},  does not \cite{marksm}. Violations of macrolocality amount to the presence of 
what a relativist would call a \textit{wormhole} \cite{lw}, a path through spacetime disallowed in a Lorentzian topology. 
General relativity allows for such paths and, in principle, they should be taken into account in a full quantum theory of gravity. In principle, in order to have traversable
wormholes, the common positive-energy conditions and some other conditions on the throats have to be satisfied. 

On the other hand, in graph-based quantum gravity states, such as in  
Loop Quantum Gravity \cite{loop}, Causets \cite{causets} or Quantum Graphity \cite{revqg}, spacetimes which are not macrolocal are very natural, and violation of macrolocality appears in the form of  non-local links.  
A first study of the physics of these non-local links was carried out in \cite{marksm,smochan}.

We propose, in the present paper, to use the framework of Quantum Graphity to provide 
 a concrete implementation of Wheeler's
quantum foam, based on the assumption that the non-local link can be used to cross from one end to the other one.

Quantum Graphity models \cite{graphity1,graphity2} are spin system toy models for emergent geometry and gravity.  They are based on quantum, dynamical graphs whose adjacency is dynamical: their edges  can be on (connected), off (disconnected),
or in a superposition of on and off. We can  interpret
the graph as pregeometry (the connectivity of
the graph tells us who is neighbouring whom). A
particular graphity model is given by such graph states
evolving under a local Ising-type Hamiltonian. The graphity model of \cite{graphity2}, for example, is a toy model for interacting matter and geometry, 
 a Bose-Hubbard model where the interactions are quantum variables.

In \cite{graphity3}, we solved the model of \cite{graphity2}  in the limit of no backreaction of the matter on the lattice, and for states with certain symmetries 
that are natural for our problem, which we called \emph{rotationally} invariant graphs.  In this case, the problem reduces to an one-dimensional Hubbard model on a lattice 
with variable vertex degree and multiple edges between the same two vertices.  
The probability density for the matter obeys
 a (discrete) differential equation closed in the classical
regime. This is a wave equation in which the vertex degree is related to the local speed of propagation of  probability. This allows
an interpretation  of the probability density of particles similar to what is usually considered in analogue gravity systems:
matter inside this analogue system sees a curved spacetime.  

We will extend these results we obtained in order to describe a quantum foam:  
instead of a classical background state (a single graph), we will  consider a state that is a superposition of many graphs. 
This amounts to  a quantum foam with a superposition of Planck scale sized non-local links. 
In our setting, the intrinsic discreteness of the graph 
sets the minimum scale.  Assuming foliability of the graph, we can define a metric distance as in \cite{graphity3}. We can then study the effect of
the \textit{quantumness} of the graph on the dispersion relations. 

Quantum Graphity models are lattice models in which the lattice becomes a quantum object. As in any lattice model, the continuum limit is obtained as in any
other lattice theory, but consider it together for all the states on the graph.

It is natural to construct graph states in which the largest contribution comes from  the 
graph with the Lorentz invariant dispersion relations.  The states with non-local links
violate macrolocality and give corrections to Lorentz invariance. 
We will construct states with a distribution of non-local links which is suppressed by their combinatorial length.
These states resemble coherent states as considered in Loop Quantum Gravity. In principle, they could be obtained as correction to the ground state due to
a non-zero temperature bath in Quantum Graphity.
The distribution depends on their density. We will then calculate the effect on the Lorentz dispersion relations in the continuum limit. 
The result is, as expected, a non-local differential equation for the evolution of the particle probability density.

It is reasonable to expect  that a non-local link will violate local Lorentz invariance. A particle can hop through the
link and behave like a superluminal particle. As we will see, the presence of all these shortcuts has an effect on the local speed of propagation of 
probability density. Also, we will find that the probability density acquires  a mass which depends on the density of non-local links. The overall dispersion relation
is thus Lorentz invariant and with a square-positive mass. However, this depends on the distribution and thus we will study two particular cases.
Using the framework of Quantum Graphity and the techniques developed in \cite{graphity3}, we will calculate the emergent mass and the constants appearing in the effective equation.

This paper is organized as follows. In section II, we summarize the  Quantum Graphity framework and the results of \cite{graphity3}. 
In section III, we show the effect of  a superposition of graphs on
the differential equation governing the time-evolution of the probability density. In section IV, we introduce our choice of the quantum state of the graph.
In section V, we analyze two particular non-local link distributions and their effect on the dispersion relations. Conclusions follow.

\label{section:model}
\section{The model}
In the following we review the model, as defined and first studied in \cite{graphity2} and the effective geometry encoded in the graph, as obtained in \cite{graphity3}.
\subsection{Bose-Hubbard model on a dynamical lattice}
In this section we will introduce briefly the model. For more a more detailed introduction we refer to the previous papers \cite{graphity2,graphity3}.

We associate a Hilbert space $\mathscr H_i$ to the degrees of freedom on the nodes a graph, with $i$ labelling the nodes. These degrees of freedom represent matter 
on the graph and thus can be, in principle, generalized to other fields.
We choose $\mathscr H_i$ to be the Hilbert space of a harmonic oscillator. We denote its creation and destruction 
operators by $b^\dagger_i$ and $b_i$ respectively, satisfying the usual bosonic commutators. 
Our $N_v$ physical systems then 
are $N_v$ bosonic modes and the total Hilbert space of such modes is given by
\begin{equation}
\mathscr H_{bosons} = \bigotimes_{i=1}^{N_v} \mathscr H_i.
\end{equation}
If the harmonic oscillators are not interacting, the total Hamiltonian is trivial:
\begin{equation}
\label{hv}
\widehat H_v = \sum_{i=1}^{N_v} \widehat H_i =- \sum_{i=1}^{N_v} \mu b^\dagger_i b_i.
\end{equation}
The Hamiltonian reads as
\begin{equation}
\widehat H = \sum_i \widehat H_i + \sum _{{\bf e}\in  I} \widehat h_{\bf e},
\label{eq:H}
\end{equation}
where $\widehat h_{\bf e}$ is a Hermitian operator on $ H_i \otimes  H_j$ representing the interaction between the system $i$ 
and the system $j$.

We introduce a primitive notion of geometry through the adjacent matrix $A$, the $N_v\times N_v$ symmetric matrix defined as
\begin{equation}
A_{ij}=\left\{{\begin{array}{ll}
1&{\mbox{if $i$ and $j$ are adjacent}}\\
0&{\mbox{otherwise}}.
\end{array}}
\right.
\end{equation}
$A$ defines a graph on $N_v$ nodes, with an edge between nodes $i$ and $j$ for every $1$ entry in the matrix.  The total Hilbert space for the graph edges is then
\begin{equation}
\mathscr H_{graph} = \bigotimes_{\bf e =1}^{N_v(N_v-1)/2} \mathscr H_{\bf e},
\end{equation}
with $\mathscr H_{\bf e}=Span\{|0\rangle,|1\rangle\}$ a qubit representing \textit{on}/\textit{off} links. Therefore, the total Hilbert space of the model is
\begin{equation}
{\mathscr H} = \mathscr H_{bosons}\otimes \mathscr H_{graph},
\end{equation}
and a basis state in $ H$ has the form
\begin{eqnarray}
|\Psi\rangle &\equiv& |\Psi^{(bosons)}\rangle\otimes|\Psi^{(graph)}\rangle \\
 &\equiv& |n_1,...,n_{N_v}\rangle\otimes |e_1,...,e_{\frac{N_v(N_v-1)}{2}}\rangle.
\end{eqnarray}
The first factor tells us how many bosons there are at every site $i$, while the second factor tells us which pairs $(i,j)$ interact.

We note that it is the
dynamics of the particles described by 
$$\widehat H_{\textrm{hop}}=- E_{hop} \sum_{i<j} A_{ij} \big(\hat a_i^\dag \hat a_i+h.c.),$$
that gives to the degree of freedom $|e\rangle$ the meaning of 
geometry and \textit{h.c.} denotes the hermitean conjugate.

The hopping amplitude is given by $t$, and therefore all the bosons have the same speed. 
Note that, for a larger Hilbert space on the links, we can have different speeds for 
the bosons.


As mentioned above, the long-term ambition of these models is to find a quantum Hamiltonian that is a spin system analogue of gravity.  
In this spirit, matter-geometry interaction is desirable as it is a central feature of general relativity.  The above dynamics can be considered as a very simple 
first step in that direction.  

In the present work, we study the model for a particular class of graphs that have been conjectured to be analogues of trapped surfaces.   
We are interested in the approximation $k\ll t$,
which can be seen as the equivalent of ignoring the backreaction of the matter on the geometry.  As in \cite{graphity3}, we will consider an Hamiltonian of the form
\begin{equation}
 \widehat H = \widehat H_v  + \widehat H_{\textrm{hop}}.
\label{eq:redH}
\end{equation}
In this case, 
the total number of particles on the graph is a conserved charge.  $\widehat H_v$ and $\widehat H_{links}$ are constants on 
fixed graphs with fixed number of particles.   The  Hamiltonian is the ordinary Bose-Hubbard model
on a fixed graph, but that graph can be unusual, with sites of varying connectivity and with more than one edge connecting two sites.
Our aim will then be to study the non-local and quantum corrections to the effective geometry which can be encoded in the graph, as shown in \cite{graphity3}.
Even on a fixed lattice, the Hubbard model is difficult to analyze, with few results in higher dimensions.  
It would seem that our problem, propagation on a lattice with connectivity which varies from site to site is also very difficult.  
Fortunately, it turns out that for our purposes it is sufficient to restrict attention to lattices with certain symmetries and then to restrict
to an effective 1+1 dimensional model.  
\subsection{Rotationally invariant graphs and the encoded geometry}
Let us present next our definition of rotationally invariant graphs, which allows to reduce the problem to a 1-dimensional Bose-Hubbard model in the single particle sector.

\begin{figure}
 \centering
 \includegraphics[scale=0.4]{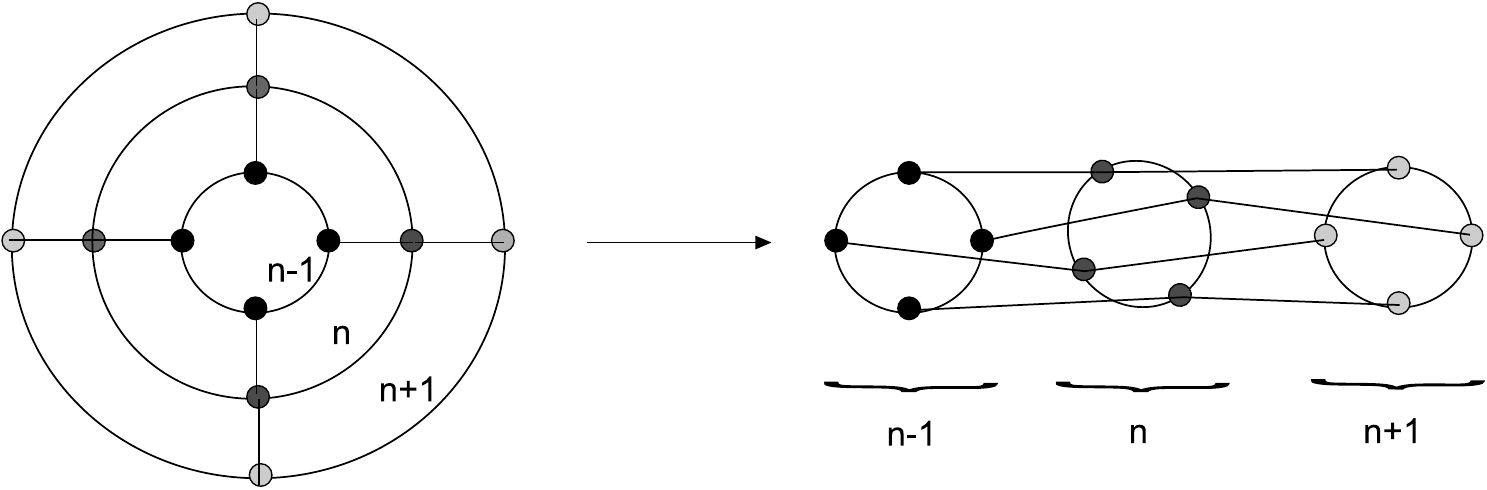}
\caption{A planar graph which is rotational invariant.}

\label{fig:rotinv1}
\end{figure}

\begin{figure}
\centering
\includegraphics[scale=0.4]{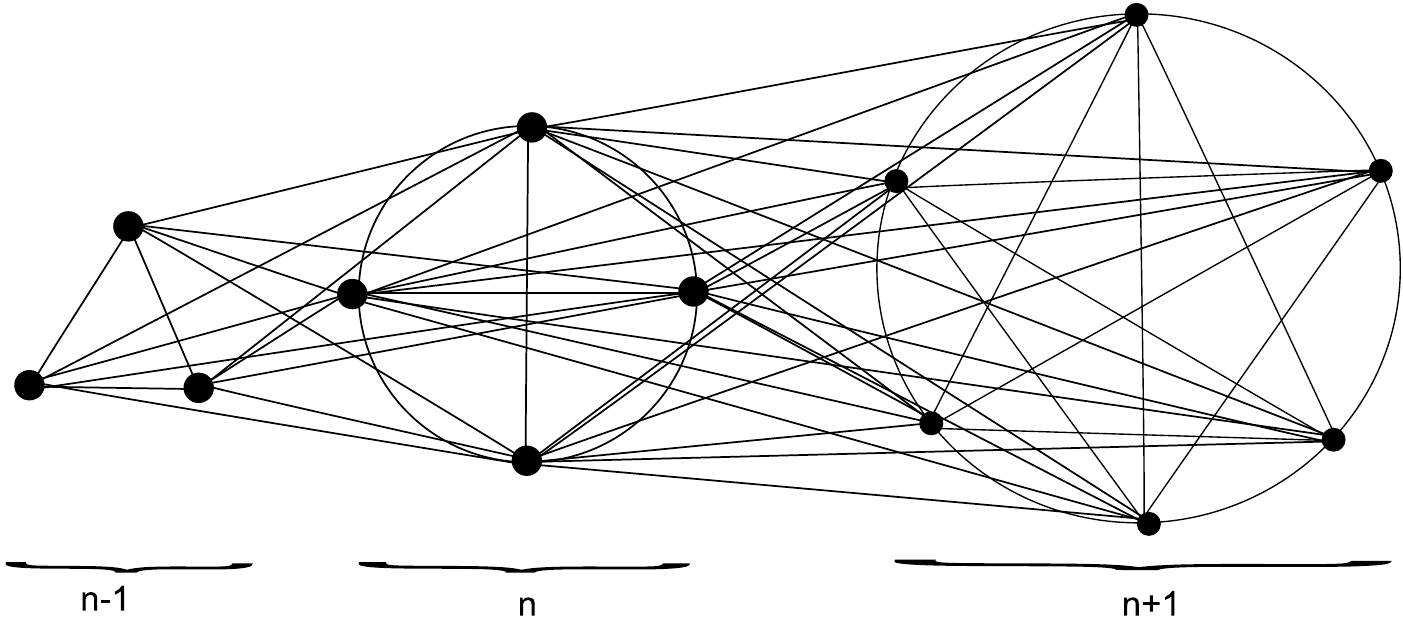}
\caption{A non-planar graph which is rotational invariant.}
\label{fig:rotinv2}
\end{figure}

A graph $G$ is called $N$-\emph{rotationally invariant} if there exists an embedding of $G$ to the plane that is invariant
by rotations of an angle $2\pi / N$. In principle, the edges of the graph, once embedded, \textit{can} be overlapping.
The main property of the rotationally invariant graphs is that groups of sub-graphs can be labelled by an integer number $i$.
These graphs can be very far from triangulations, as the rotationally invariant graphs in Fig. \ref{fig:rotinv1} and \ref{fig:rotinv2} show.

These graphs can be labelled by a set of two integer coordinates, $(n, \theta_n)$, where $n$ labels a set of nodes, while $\theta_n$ is
a coordinate internal to the subgraph. For convenience we will drop, since now on, the sub-index $n$ in the $\theta$ coordinates. 

We can make use of the coordinates $(n, \theta)$ in order to write
the Hamiltonian defined by a rotationally invariant graph as 
\begin{align}
H_{\textrm{rot}}&=-\sum_{\theta=0}^{N-1}\sum_{n,n'} A_{nn'}b_{n\theta}^\dagger b_{n'\theta} +h.c.\nonumber \\
&-\sum_{\theta=0}^{N-1}\sum_{\varphi=1}^{N-1}\sum_{n,n'} B_{n,n'}^{(\varphi)}b_{n\theta}^\dagger b_{n' \theta+\varphi}+h.c., 
\end{align}
where $b_{n,\theta}^\dagger$ ($b_{n,\theta}$) is the creation (annihilation) operator at the vertex $(n,\theta)$,
$A_{nn'}$ is the adjacency matrix of the graph and $B_{n,n'}^{(\varphi)}$
is the adjacency matrix of two angular sectors at an angular distance $\varphi$ in units of $2\pi / N$.
 
Let us introduce the rotation operator $\widehat M$ defined by
\begin{align}
\widehat M b_{n,\theta} = b_{r,\theta +1} \widehat M \, \nonumber \\
\widehat M b_{n,\theta}^\dagger = b_{r,\theta +1}^\dagger \widehat M \, .
\label{eq:L}
\end{align}
The effect of the operator $\hat M$ is particularly easy to understand in the single particle case:
\be
\widehat M \ket{n,\theta} = \widehat M b_{n,\theta}^\dagger \ket{0} = b_{n,\theta +1}^\dagger \widehat M \ket{0}  = \ket{n,\theta+1}\, ,
\ee
where we have assumed that the vacuum is invariant under a rotation $\widehat M\ket{0}=\ket{0}$.
Note that $\widehat M$ is unitary 
and its application $N$ times gives the identity, $\widehat M^N=\id$. 
This implies that its eigenvalues are integer multiples of $2\pi / M$.

Another interesting property of $\widehat M$ is that commutes both with the rotationally invariant 
Hamiltonians and with the number operator $\widehat N_p$,
\be
	[\widehat H_{\textrm{rot}},\widehat M]=[\widehat N_p,\widehat M]=[\widehat H_{\textrm{rot}},\widehat N_p]=0 \, .
\ee
Therefore $\widehat H_\textrm{rot}$, $\widehat N_p$, and $\widehat M$ form a complete set of commuting observables and
the Hamiltonian is diagonal in blocks of constant $\widehat M$ and $\widehat N_p$.
In this sector of the Hamiltonian, we can reduce the Hamiltonian to:
\be
\widehat H_0 = \sum_{n=0}^{L-1} f_{n,n+1}\left(\ket{n}\bra{n+1}+\ket{n+1}\bra{n}\right) + \sum_n \mu_n \proj{n}\, , 
\label{eq:Hamiltonian-BH-single-particle}
\ee
with $f_{n,n+1}$ depending on the degree of the graph and $n$ being the label of the shells we are reducing with the rotational symmetry and $L$ the total size of the one-dimensional lattice.
\subsection{Restriction of the time-dependent Schr\"odinger equation to the set of classical states}

Since we want to study the dynamics of a single particle on a fixed graph, 
it is only necessary to consider the single particle sector.
The one dimensional Bose-Hubbard model for a single particle reads as in (\ref{eq:Hamiltonian-BH-single-particle}),
where $f_{n,n+1}$ are the tunneling coefficients between sites $n$ and $n+1$,
$\mu_n$ is the chemical potential at the site $n$, {and $M$ is the size of the lattice.

%

In this setup, let us introduce the convex set of classical states $\mathcal{M}_C$, 
parameterized as
\be
\widehat \rho(t)\equiv\widehat \rho\big(\Psi(t)\big) = \sum_{n=0}^{L-1} \Psi_n \proj{n}\, ,
\ee
where $\Psi_n$ is the probability of finding the particle at the site $n$.
The states in $\mathcal{M}_C$ are classical because the uncertainty in the position is classical, that is,
they represent a particle with an unknown but well-defined position.

Since our particle is under the effect of a noisy environment,
its density matrix is going to be constantly dephased by the interaction between the particle and its reservoir. For a more detailed discussion
about this procedure and the connection with the physics of decoherence, we refer to \cite{graphity3}.
The dephased state in the position eigenbasis that best approximates $\rho(t+\Delta t)$ can be easily determined by computing
the double commutator of the previous equation, which was shown to lead to a closed equation in \cite{graphity3}. 
It obeys the evolution
\begin{eqnarray}
\frac{\hbar^2}{2}\partial_t^2 \Psi_n(t) =& f_{n-1,n}^2 \left(\Psi_{n+1}(t)+\Psi_{n-1}(t)-2 \Psi_{n}(t)\right) \nonumber  \\
&+\left(f_{n+1,n}^2-f_{n-1,n}^2\right)\left( \Psi_{n+1}(t)-\Psi_{n}(t)\right) \, .\nonumber
\label{eq:EOM}
\end{eqnarray}
This equation becomes a wave equation in the continuum,
\begin{equation}
\partial_t ^2 \Psi(x,t)-\partial_x \left(c^2(x)\partial_x \Psi(x,t)\right)=0\, ,
\label{eq:continuous-EOM}
\end{equation}
where 
\begin{equation}
 \frac{1}{c(x)}=\sqrt{ \frac{\hbar^2}{2f^2(x)E_{\textrm{hop}}^2} } = \frac{\hbar}{E_{\textrm{hop}}\sqrt{2f^2(x)}}\, ,
\label{eq:speed-of-light}
\end{equation}
and $\Psi(x,t)$ and $f(x)$ are the continuous limit functions of $\Psi_{n}(t)$ and $f_{n,n-1}$ respectively. Eqn (\ref{eq:continuous-EOM}) is the equation of motion for
a scalar field with a space-dependent refraction index. As it is well known, this equation in higher dimension is connected with the Gordon metric. In fact, to the refraction index
it is possible to encode a space-time geometry with spatial curvature and no extrinsic curvature, i.e. a preferred direction of time. The time direction is the same of the quantum mechanical underlying model.
This equation is the starting point for what we will do in the following. However, let us first recall how the continuum limit is performed.

\subsection{Dispersion relation and continuum limit}
\label{section:dispersion}
Let us consider in more detail the translationally invariant case in which $f_{n-1,n}=f$ and $\mu_n=\mu$ for all $n$.
In this case, the continuous wave equation \eqref{eq:continuous-EOM} becomes
\be
\partial_t ^2 \Psi(x,t)-c^2\partial_x^2  \Psi(x,t)=0\, ,
\ee
where $c$ is the speed of propagation. 

Let us recall how this limit was performed in \cite{graphity3}. Let us first introduce a discrete Fourier transform in the spatial coordinate and a continuous Fourier transform in the temporal coordinate, 
given by
\be
\Psi_n(t) = \frac{1}{\sqrt{L}}\sum_{k=0}^{L-1} \tilde \Psi_k(t) e^{-\iu \frac{2\pi}{L} nk}\, ,
\label{discf}
\ee 
and $\tilde \Psi_k(t)= A e^{\iu \omega_k t}+ B e^{-\iu \omega_k t}$.
After a straightforward calculation, we find that the relation between $\omega_k$ and $k$ is given by
\begin{equation}
\omega_k\ c  = \sqrt{2}\ \sqrt{1-\cos\left(\frac{2\pi}{L}k\right)}.
\label{discdisp}
\end{equation}
Now we can rescale $\omega_k\rightarrow \tilde\omega_k/L$ (or equivalently $c$) and find that 
\begin{equation}
\tilde \omega_k c=  L\sqrt{2} \sqrt{1-\cos\left(\frac{2\pi}{L}k\right)},
\label{contdisp}
\end{equation}
and, therefore,
$$\lim_{L\rightarrow \infty} \tilde \omega_k(L)\approx 2 \pi \frac{k}{c}. $$
That is, only  modes that are slow with respect to the time scale set by $c$ see the continuum. Note that by rescaling the speed of propagation $c$, the continuum limit can be obtained by a double scaling limit,
$ E_{\textrm{hop}}\rightarrow E_{\textrm{hop}}/L$ and $L\rightarrow\infty$ for lattice size $L$. In this limit, the probability 
density has a Lorentz invariant dispersion relation.

\section{A non-local state distribution}
In this section, we show the effect of having a quantum superposition of graph in (\ref{born})  on the equation (\ref{eq:EOM}).
\subsection{The effect of a quantum superposition of graphs}
In order to do the explicit calculation, we will modify the Bose-Hubbard interaction. Let us consider a one-dimensional
Bose-Hubbard of the form,
\begin{equation}
 \widehat H = \sum_{i} A_{i,i-1} (\hat a_i^\dag \hat a_i + h.c.)
\end{equation}
and then consider its generalization, from $A_{i,j}=\delta_{j,i-1}+\delta_{j,i+1}$, to $\widehat A_{i,j}=\widehat N_{ij}$, with
$\widehat N_{ij}=\hat b^\dagger_{ij} \hat b_{ij}$  and $\hat b_{ij}$,$\hat b^\dagger_{ij}$ the ladder operators on the Hilbert space of the link $ij$. $\widehat N_{ij}$ is then the number operator on the Hilbert space of the graph, as usually considered in Quantum Graphity. 
This allows, instead of using fixed \textit{classical} graphs, fixed \textit{quantum} graphs, where the state $|\psi_{graph}\rangle$ is superposition
of different graphs. The full quantum hamiltonian for the system is, as usual, on an Hilbert space of the form
$$|\psi_{total}\rangle=Span\{|\psi_{graph}\rangle \otimes |\psi_{bosons}\rangle\}. $$

Using this, we want now to repeat the same calculation we performed in the previous paper, i.e. compute:
\begin{equation}
 \partial_t^2 \psi_z(t) = -i\ Tr\{[\widehat H,[\widehat H,\widehat \rho(t)]] \widehat N^\prime_z\},
\end{equation}
with $\psi_n=\langle\widehat N^\prime_n \rangle$, $\widehat N^\prime_n$ number operator on the bosons defined on the node $n$, and $\widehat \rho$ the density matrix on the total system.
Let us assume that the graph is not dynamical. We will also to use the Born approximation, that is,
\begin{equation}
\widehat \rho(t)\approx \widehat \rho_{g} \otimes \widehat \rho_b(t),
\label{born} 
\end{equation}
with $\widehat \rho_g$ the density matrix of the graph and $\widehat \rho_b(t)$ the density matrix of the bosons.
This approximation allows us to consider a particle disentangled enough from the graph to be ``followed'' using the equation (\ref{eq:continuous-EOM}). 
It is also a physical requirement, 
which accounts for the existence of the particle on its own. In general, we expect that at long times the full hamiltonian thermalizes to a specific graph, 
depending on the parameter of the Hamiltonian which defines the metastable state. Later on, we will rescale the coupling constant of the hopping Hamiltonian in order to 
obtain the continuum limit. Thus, one could think that this rescaling affects the
state of the graph at infinity. However, the hopping of the bosons allows the graph to thermalize, as it has been shown in \cite{graphity2}. 
Rescaling this constant, just changes the time it takes for the system to thermalize, but not
the asymptotic state of the graph.  
As a matter of fact, we do not know yet a Hamiltonian which gives a specific graph state asymptotically. However, the results of \cite{florian} in two dimensions and those 
of \cite{graphity1}, support the conjecture
that, in general, such a Hamiltonian exists. For the time being, it is fair to say that the ground state of Quantum Graphity coupled to a thermal bath are rotational invariant 
graphs \cite{konopka}. Thus, these graphs can at least be generated by a known effectively 2d-dimensional model.

Based on these considerations, we conjecture the following graph state, $|\psi_{graph}\rangle=|\psi_{cl}\rangle +|\psi_{nl}\rangle$ with
$\langle \psi_{cl}|\psi_{nl}\rangle=0$. $|\psi_{nl}\rangle$ is a correction to the classical graph state $|\psi_{cl}\rangle$ considered in \cite{graphity3} that we will discuss (and construct) in 
the next section. For the time being, let us consider the effect of this correction on eqn. (\ref{eq:EOM}).
 We have $\widehat \rho_g=|\psi_{graph}\rangle\langle \psi_{graph}|$.  Thus:
\begin{equation}
 \widehat \rho_{g}=|\psi_{cl}\rangle\langle\psi_{cl}|+ |\psi_{nl}\rangle\langle\psi_{nl}|+(|\psi_{nl}\rangle\langle\psi_{cl}|+|\psi_{cl}\rangle\langle\psi_{nl}|).
\end{equation}
Let us now evaluate these traces. A straightforward calculation shows that,
\begin{eqnarray}
-\frac{E_{hop}^2}{\hbar^2} \partial_t^2 \psi_n &=&  Tr\ \{ (\widehat H^2 \widehat \rho + \widehat \rho \widehat H^2-2 \widehat H \widehat \rho  \widehat H) \widehat N_z\} \nonumber \\
&=& 2 \sum_{ij,mn}\big[ Tr\{\widehat A_{ij} \widehat A_{mn} \widehat \rho_g\} Tr\{\hat a_i^\dag \hat a_j \hat a_m^\dag \hat a_n \widehat \rho_b \widehat N_z\} \nonumber \\
&-& Tr\{\widehat A_{ij}  \widehat \rho_g \widehat A_{mn} \} Tr\{\hat a_i^\dag \hat a_j \widehat \rho_b \hat a_m^\dag \hat a_n \widehat N_z\} \big]. 
\end{eqnarray}
We now substitute the equation for $\widehat \rho_g$, and obtain:
$$ -\frac{E_{hop}^2}{\hbar^2} \partial_t^2 \psi_n = \tilde \triangle \psi_n(t) +  C_n(t), $$
with $\tilde \triangle \psi_n(t) $ is the discrete second derivative and $C_n(t)$ is:
\begin{eqnarray}
C_n(t)&=2\sum_{ij,mn} \big[P_{ijmn} Tr\{\hat a_i^\dag \hat a_j \hat a_m^\dag \hat a_n  \widehat \rho_b \widehat N_z\}\nonumber \\
&- Q_{ij}Q_{mn} Tr\{\hat a_i^\dag \hat a_j \widehat \rho_b \hat a_m^\dag \hat a_n \widehat N_z\}\big],
\end{eqnarray}
with:
$$ P_{ijmn}=\langle\psi_{nl}|\widehat A_{ij} \widehat A_{mn}|\psi_{nl}\rangle,$$
$$ Q_{ij}=\langle\psi_{nl}|\widehat A_{ij}|\psi_{nl}\rangle,$$
where we used the orthogonality condition $\langle \psi_{nl}|\psi_{cl}\rangle=0$.

Our task now is to evaluate these two quantities on different classes of interesting states.
\section{The choice of the quantum state for the graph} 
\begin{figure}
\centering
\includegraphics[scale=0.4]{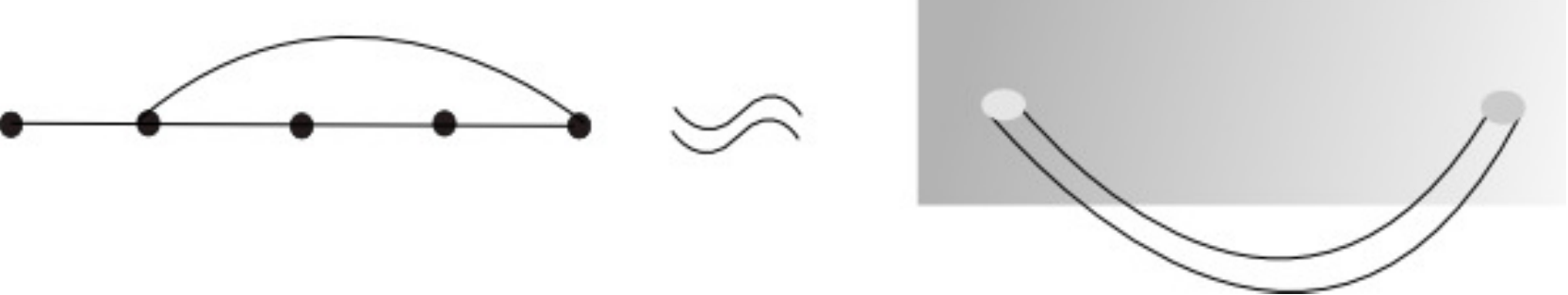}
\caption{The intuitive picture of non-local links inserted in the graph.}
\label{Fig1-wh}
\end{figure}

Let us now introduce the states on which we will evaluate the quantities defined in the previous section, $P_{ijmn}$ and $Q_{ij}$.
Motivated by the fact that we can reduce using the translationally symmetric graphs to one line, we will restrict our attention to a one-dimensional lattices.
These states will resemble coherent states as considered in Loop Quantum Gravity. In principle, they could be obtained as correction to the ground state due to
a non-zero temperature bath in Quantum Graphity.

Let us consider first a metric on the classical graph, with $d(i,j)$ the distance between the nodes of the classical 
graph $|\psi_{cl}\rangle$, with all the ordinary properties of distances. On a one-dimensional line this distance could be, for instance, given by $|i-j|$.
Let us then construct states with non-local links on top. We want to penalize states with too long non-local links. We then introduce a factor $\rho(i,j)$, which depends on a distance $d(i,j)$ evaluated on the base graph, 
assuming that $d(i,j)\geq0$, and a parameter $l$ describing how 
non-local the links are w.r.t. the length of the graph. Then we define the operator:
\begin{equation}
\widehat T_l= \sum_{i<j} \rho(i,j)\ \hat a_{ij}^\dag, 
\end{equation}
with
\begin{equation}
\sum_{i<j} \rho(i,j)^2=1,  
\end{equation}
which ensures that $\rho(i,j)^2$ can be interpreted as a classical probability distribution.
When applied to $|\psi_{cl}\rangle$ this operator generates a superposition of all the possible non-local links which can be created on $|\psi_{cl}\rangle$, with a factor that
with the distance of the links,
\begin{equation}
|\psi^1_{nl}\rangle=\widehat T_l |\psi_{cl}\rangle, 
\end{equation}
and we can imagine to apply this operator several times to create more non-local links,
\begin{equation}
|\psi^R_{nl}\rangle=\widehat \frac{T_l^R}{R!} |\psi_{cl}\rangle.
\end{equation}
The meaning to give to $l$ is thus that of a cut-off in the length of these non-local links.  Note that we can bias the number of links on which we want to peak the 
quantum non-local state the same way,
\begin{equation}
\widehat \mathscr T^K_l=\sum_{s=1}^{\infty} \frac{K^s}{s!} \widehat T^s_l = e^{K \widehat T_l}-1.
\end{equation}
We see then that we can write the quantum state for the graph in the convenient form:
\begin{equation}
|\psi_{nl}\rangle=\big[1+ e^{K \widehat T_l}\big]|\psi_{cl}\rangle.
\end{equation}
This state depends explicitly on two parameters, $l$ and $K$, and on the classical graph together with its distance.
On this state we now want to evaluate:
\begin{equation}
 P_{ijmn}=\langle \psi_{nl}|\widehat A_{ij} \widehat A_{mn}|\psi_{nl}\rangle=\langle \psi_{cl}|\widehat \mathscr T_l^{K \dag } \widehat A_{ij} \widehat A_{mn} \widehat \mathscr T_l^{K} |\psi_{cl}\rangle,
\end{equation}
and
\begin{equation}
 Q_{ij}=\langle \psi_{nl}|\widehat A_{ij} |\psi_{nl}\rangle=\langle \psi_{cl}|\widehat \mathscr T_l^{K \dag } \widehat A_{ij} \widehat \mathscr T_l^{K} |\psi_{cl}\rangle.
\end{equation}

Let us  then consider first the average. We note that, since $\widehat A_{ij}$ acts like a projector, and
states with different powers of the $\widehat T_l$ operators are orthogonal, we can write:
\begin{equation}
Q_{ij}=\sum_{s=1}^\infty \frac{K^{2s}}{{s!}^2} \langle \psi_{cl}|T^{\dag s}_l \widehat A_{ij} T^{s}_l |\psi_{cl}\rangle .
\end{equation}

To clarify the idea, let us consider the case in which we add just a link. In this case, the state is the sum over all possible
links which can be created, with a factor $\rho^2(i,j)$.This link can be created in one way only, and so $\widehat A_{ij}$ projects
on the only state which can be non-zero. A very straightforward calculation shows that
\begin{equation}
 \langle \psi_{cl}|T^{\dag 1}_l \widehat A_{ij} T^{1}_l |\psi_{cl}\rangle = 2\ \rho^2(i,j).
\end{equation}
For the higher order term, we instead have:
\begin{equation}
 \langle \psi_{cl}|T^{\dag s}_l \widehat A_{ij} T^{s}_l |\psi_{cl}\rangle = \rho^2(i,j) \sum_{i_1,j_1,\cdots,i_{s-1},j_{s-1}} \prod_{l=1}^{s-1} \rho^2(i_l,j_l).
\end{equation}
It is easy to see that
\begin{equation}
\sum_{i_1,j_1,\cdots,i_{s-1},j_{s-1}} \prod_{l=1}^{s} \rho^2(i_l,j_l)\approx 2^s\ s\ (l\ L)^s,
\label{eq:dist_const}
\end{equation}
due to the fact that the integration is over the line, while the distribution has an extension of \textit{circa} $l$ combinatorial points. The factor $2^s$ comes from the fact that
there are 2 points we are summing over and the $s$ factor from the $s$ sums appearing in $T^s_l$.
Thus, we can write:
\begin{equation}
 \langle \psi_{cl}|T^{\dag s}_l \widehat A_{ij} T^{s}_l |\psi_{cl}\rangle =c_s\ \rho^2(i,j)\ 2^s\ s\ (L\ l)^{s-1}.
 \label{evalqij}
\end{equation}
In principle, given a distribution, we can calculate this factor from eq. (\ref{eq:dist_const}). We will calculate these factors later for two
particular distributions. Plugging eqn. (\ref{evalqij}) into $Q_{ij}$, we obtain
\begin{equation}
 Q_{ij}=\sum_{s=1}^\infty \frac{K^{2s}}{{s!}^2}\ 2^s\ s\  (L\ l)^{s-1} \rho^2(i,j) c_s =  \rho^2(i,j) R(K,l\ L),
\end{equation}
with:
\begin{equation}
 R(K,l\ L)= \sum_{s=1}^\infty \frac{K^{2s}}{{s!}^2}\ c_s\ 2^s\ s\ (l\ L)^{s-1},
\end{equation}
and, therefore,
\begin{equation}
 Q_{ij}Q_{mn}= \rho^2(i,j) \rho^2(m,n) R(K,l\ L)^2.
\end{equation}
We can, in fact, do an analogous calculation for $P_{ij mn}$ and find that:
\begin{equation}
 P_{ijmn} = \rho^2(i,j) \rho^2(m,n) L(K,l\ L),
\end{equation}
with:
\begin{equation}
 L(K,l\ L)= \sum_{s=1}^\infty \frac{K^{2s}}{{s!}^2} c_s (l\ L)^{s-2}\ 2^s\ s.
\end{equation}

Going back to the original problem, we find  that the correction to the discrete Lorentz equation is:
\begin{eqnarray}
 C_z =& 2\sum_{ij,mn} \rho^2(i,j) \rho^2(m,n) \big[ L(K,l\ L) Tr\{\hat a_i^\dag \hat a_j \hat a_m^\dag \hat a_n  \widehat \rho_b \widehat N_z\} \nonumber \\
&-R(K,l\ L)^2 Tr\{\hat a_i^\dag \hat a_j \widehat \rho_b \hat a_m^\dag \hat a_n \widehat N_z\}\big].
\end{eqnarray}
If we define:
\begin{equation}
 S(K,l\ L)= (l\ L)\ R(K,l\ L) = (l\ L)^2\ L(K,l\ L),
\end{equation}
with $S(K,l\ L)=\sum_{s=1}^\infty \frac{K^{2s}}{{s!}^2}\ c_s\ (l\ L)^{s}\ 2^s\ s$, then we obtain:
\begin{equation}
C_z(t)=2\sum_{ij,mn} \rho^2(i,j) \rho^2(m,n) S(K,l\ L)\big[Tr\{\hat a_i^\dag \hat a_j \hat a_m^\dag \hat a_n  \widehat \rho_b\widehat N_z\}- S(K,l\ L) Tr\{\hat a_i^\dag \hat a_j \widehat \rho_b \hat a_m^\dag \hat a_n \widehat N_z\}\big].
\label{correction}
\end{equation}
We see that the function $S(K,l\ L)$ depends, as a matter of fact, on $\xi=K\sqrt{l\ L}$,
$S(K,l\ L)\equiv S(\xi)=\sum_{s=1}^\infty\ c_s\ [\frac{\xi^{s}}{{s!}}]^2\ 2^s\ s$.
A plot of the function $S(K,lL)$ can be found in Fig. \ref{FigF}.
\begin{figure}
\centering 
 \includegraphics[scale=0.4]{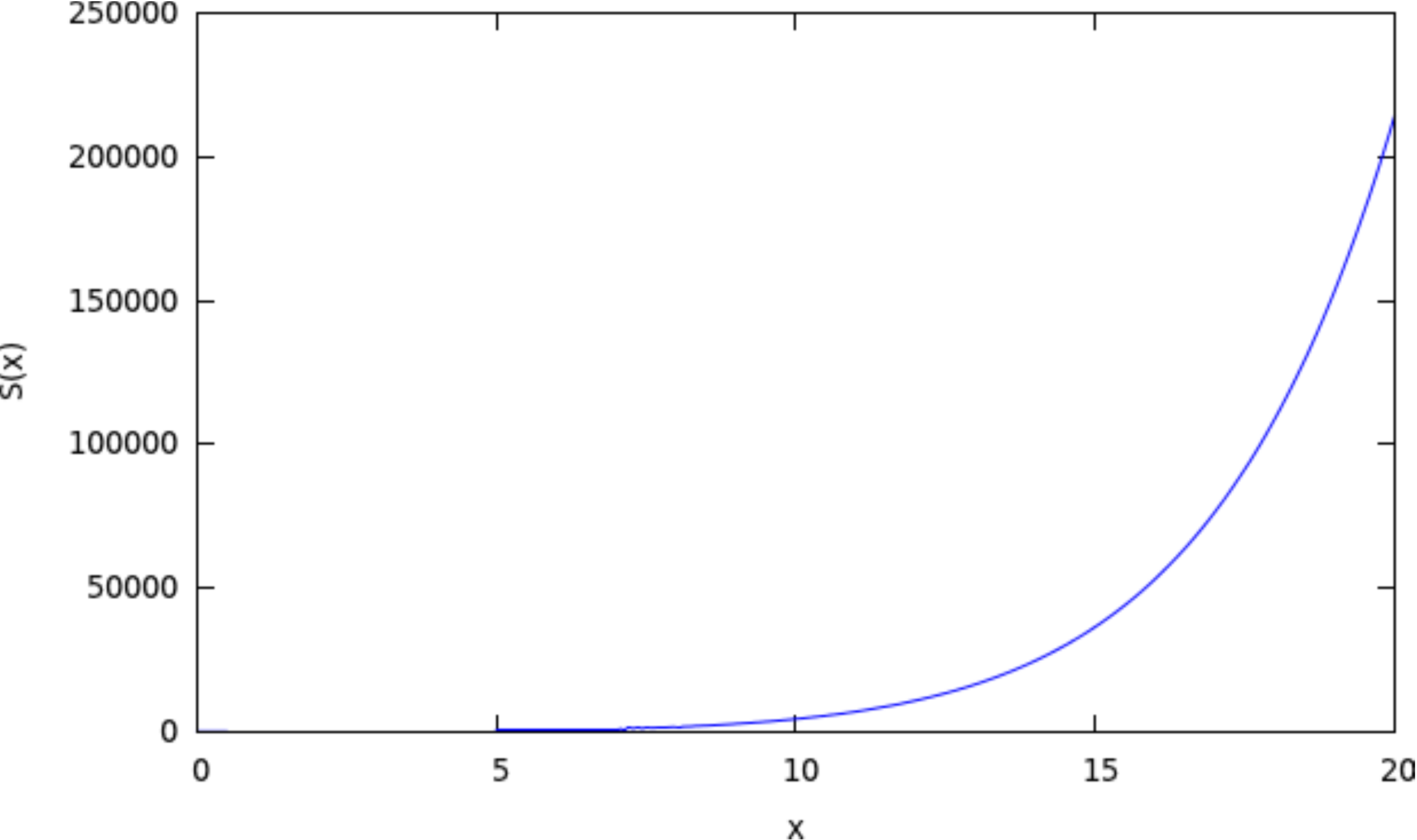}
\caption{A plot of the function $S(x)$ which appears in eq. (\ref{correction}).}
 \label{FigF}
\end{figure}
The traces can be evaluated, as done in (\cite{graphity3}), and the result is:
\begin{equation}
C_z(t)=2\frac{1}{(l\ L)^2}\sum_{j} \rho^4(z,j) S(K,l\ L)(\psi_z-S(K,l\ L)\psi_j).
\end{equation}

Some comments are now in order. First of all, note that the equation has the shape of a second derivative. To understand this, we can look at a term of the form $\sum_{|k|\geq 2} J(k)(\psi_z - \psi_{z+k}) $.
This term can be written as:
$$\sum_k J(k)\cdots=-J(2)\big(\psi_{z+2}-2\psi_z+\psi_{z-2}\big)-J(3)\big(\psi_{z+3}-2\psi_z+\psi_{z-3}\big)-J(4)\cdots. $$
This is a sum of discrete second derivatives with a non-local mass, 
$$J(k)\big(\psi_{z+k}-2\psi_z+\psi_{z-k}\big)=-J(k) \big(\psi_{z+k-1}+\psi_{z-k+1}\big)-J(k)\sum_{i=2}^{k-1}\big(\psi_{z+i+1}-2\psi_{z+i}+\psi_{z+i-1}\big),$$
so we expect, in the end, to obtain a mass term out of this equation and, when we will have rearranged all the terms, we will.

Note that, for the case $c_s=1$, $S(\xi)=1$ for $\xi=0.903$, and so $K=\frac{0.903}{\sqrt{l\ L}} $. We then see that $K^2$ plays the role of the density of non-local links per units of $l\ L$.

To end this section, we have to calculate the norm of this state. This can be written as:
\begin{equation}
 |\langle\psi_{nl}|\psi_{nl}\rangle|=\sqrt{1+ \langle\psi_{cl}|e^{K \widehat T^\dag_l}e^{K \widehat T_l} |\psi_{cl}\rangle^2 +2 \Re\{\langle \psi_{cl}| e^{K \widehat T_l}|\psi_{cl}\rangle \}  },
\end{equation}
which reads,
\begin{equation}
|\langle\psi_{nl}|\psi_{nl}\rangle|=\sqrt{1+ \big(\langle\psi_{cl}|e^{K \widehat T^\dag_l}e^{K \widehat T_l} |\psi_{cl}\rangle^2-1\big)}\}
\end{equation}
and substituting for the $\mathscr T$ operators, we finally find
\begin{equation}
\mathscr N=|\langle\psi_{nl}|\psi_{nl}\rangle|=\sqrt{1+ \big(\sum_{s=1} \frac{K^{2s}}{(s!)^2} \sum_{i<j} \rho(i,j)^2\big)^2}=\sqrt{1+ \big(\sum_{s=1} \frac{K^{2s}}{(s!)^2} 2^s s\big)^2}=\sqrt{1+ S^2(K,l\ L)}.
\end{equation}
We can thus normalize the graph state by dividing by a factor of $\mathscr N$.

\section{The modified dispersion relation due to disordered locality}
\textit{The general case.} We will now discuss the continuum limit. As we have seen, the continuum limit is obtained by rescaling $E_{hop}\rightarrow \tilde E_{hop}/L$ and then 
sending $L\rightarrow \infty$.  Please note that $E_{hop}$ appears whenever we hop with a particle, so in these calculations it appears everywhere but in the $\partial_t^2$ term. 
In order to perform the continuum limit, first we have to make sense of the quantity $(\psi_z-S(K,l\ L)\psi_j)$ 
at least for the \textit{flat} case, which we know correspond to Lorentz from \cite{graphity3}. 
We can add and subtract,
\begin{eqnarray}
&(S(K,l\ L)-1)\psi_z+(S(K,l\ L)\psi_z-S(K,l\ L)\psi_j)=\nonumber \\
& = (S(K,l\ L)-1)\psi_z+S(K,l\ L)(\psi_z-\psi_{z-1}+\psi_{z-1}+\psi_{z-2}-\cdots-\psi_j). \nonumber 
\end{eqnarray}
In the continuum limit this becomes
$(S(K,l\ L)-1)\psi(z,t)+S(K,l\ L)\int_{j}^z \partial_\xi \psi(\xi,t) d\xi$, and thus $C_z(t)$ reads:
\begin{equation}
C_z(t)=\int_L dx\  \rho^4(z,x) [\frac{(S(K,l\ L)-1)S(K,l\ L)}{(l\ L)^2}\psi(z,t)+\frac{S^2(K,l\ L)}{(l\ L)^2}\int_{x}^z \partial_\xi \psi(\xi,t) d\xi], 
\end{equation}
which is:
\begin{equation}
C_z(t)=\psi(z,t)\int_L dx\ \rho^4(z,x) \frac{(S(K,l\ L)-1)S(K,l\ L)}{(l\ L)^2}+\frac{S^2(K,l\ L)}{(l\ L)^2} \int_L \rho^4(z,x) \int_{x}^z \partial_\xi \psi(\xi,t)\ d\xi\ dx.
\end{equation}
This can be written as:
$$C_z(t)=\psi(z,t) F(K,l\ L)+ O(K,l\ L) \int_L \rho^4(z,x) \int_{x}^z \partial_\xi \psi(\xi,t)\ d\xi\ dx,$$
with $F(K,l\ L)=\int_L dx \rho^4(z,x) \frac{(S(K,l\ L)-1)S(K,l\ L)}{(l\ L)^2}$,$O(K,l\ L)=\frac{S^2(K,l\ L)}{(l\ L)^2}$. 
Please note here that these steps have been performed naively, though we have an explicit dependence on $L$ in $S$. It is important to point out that
the only way to keep the function $S(l\ L,K)$ finite is to rescale the quantity $K^2 l \approx \frac{\tilde K^2 \tilde l}{L}$. 
To keep the discussion simple,
let us discuss this point at the end of the section. $L$ is the combinatorial length of the 1-d lattice we are considering, and over which $\psi(x,t)$ is defined. Thus the equation of motion for the flat case is given, in the continuum, by:
$$[\partial_t^2-c^2\big(1+S^2(K,l\ L)\big) \partial_z^2 - F(K,l\ L)]\psi(z,t)= O(K,l\ L) \int_L \rho^4(z,x) \int_{x}^z \partial_\xi \psi(\xi,t)\ d\xi\ dx,$$ 
which is an integro-differential equation for the field integrated over the line, which shows the strong non-local character of the equation. 

We note that there is a contribution to the speed of propagation of the signal, due
to the fact that particle can hop on many more graphs than the single classical one. 
This factor contributes with a $c^2 S^2(K,l\ L)$ added to the effective speed $c^2$. Let us stress that this contribution
is merely due to the fact that there are many more graphs in the superposition, and not due to the fact that the particle can hop further: 
this is kept track of in the $C_z(t)$ term of the equation.
Also, we see that $F(l,K)$ becomes a mass, due to non-locality, while on the r.h.s. there a new term appears. We can further reduce the equation by evaluating the integrals. It is clear
that in order to have a finite result, which is physically expected, we have to rescale at this point only $l\approx \tilde l/L$, keeping $K^2$ independent from $L$. 
Said this, we see
that the distribution itself, when is well chosen, becomes a $\delta$ function and therefore the models becomes local again. 

Let us now calculate the terms at the leading order in $1/L$, since that is what we are interested in.
The discrete differential equation becomes: 
\begin{equation}
[\partial_t^2-c^2\big(1+S^2(K,l\ L)\big) \tilde \partial_z^2 - \tilde F(K,l\ L)]\psi_z(t)= -O(K,l\ L) \sum_{x=0}^{L} \rho^4(z,x)\psi_z(t),
\end{equation}
where $\tilde \partial_z^2$ is the discrete spatial second derivative.
Using now (\ref{discf}), we see that the dispersion relation for the field becomes:
$$ \omega_k c\big(1+S^2(K,l\ L)\big)= \sqrt{2}\sqrt{1-\cos\big(\frac{2\pi}{L}k \big)+\tilde F(K,l\ L)+\tilde \rho^4(k) O(K,l\ L)} $$
Please note that with this rescaling of $K$, we have that $S(K,l\ L)$ can be expanded in even powers of $1/L$:
$$S(K,l\ L)=2 \frac{\tilde K^2 \ l\ L }{L^2}+8 \frac{ \tilde K^2 (l\ L)^2 }{L^4}+\cdots. $$
Thus, we see that the superluminal effect, which is, the factor $1+S^2(K,l\ L)$, becomes one in the limit $L\rightarrow\infty$; also, in the same limit, only the part quadratic in $K$
survives. At this point the equation would become, in the continuum:
\begin{equation}
[\partial_t^2-c^2 \partial_z^2 - \tilde F(K,l\ L)]\psi(z,t)= -O(K,l\ L) \int_L \rho^4(z,x)\psi(z,t)\ dz 
\end{equation}
with $\tilde F(k,l\ L)=F(K,l\ L)+O(K,l\ L)\int_L \rho^4(z,x)\ dx$. 
Note that, while $\tilde F$ might seem to be dependent on the
point $z$, being $\tilde F$ dependent on $z-x$ and integrated over $x$, it is indeed independent from it. 
In particular, if we define $l\ L\equiv \xi$, in the limit $L\rightarrow \infty$ and with the rescaling of $K$ and $l$, $ S(K,l\ L)  \rightarrow 2 \tilde K^2 \xi$.
We see now that the only way to obtain the continuum dispersion relation by rescaling $c\rightarrow \tilde c/L$, as done for the single-graph state, is to rescale also $K$, with $K\rightarrow \tilde K/L$.

Just as an exercise, we can insert a trivial spatially-constant solution, which then becomes of the form 
$\partial_t^2 \psi(t)=R(K,l\ L) \psi(t) $.
where $R(K,l\ L)=F(K,l\ L)+2\ O(K,l\ L)\ \int_L\ \rho^4(z,x)\ dx$. Note that this quantity is always positive, so constant solutions are stable. Let us try to find a generic solution, instead. 
Let's do it for the equation:
\begin{equation}
[\partial_t^2- \tilde c^2\partial_x^2 +\tilde c^2 q]\psi(x,t)=-\tilde c^2 \int_L \sigma (z,x) \psi(y,t)\ dy . 
\end{equation}
Since the equation is linear in the field $\psi$, we can solve it by means of a Fourier transform. We then look at the dispersion relation for the function $\psi(x,t)$, with $q$ and $P$ generic functions. We can do it by Fourier transform. In this case, the integral on the right, being a convolution,
becomes just the product of the Fourier transform of the single functions. Thus we have:
$$ -\omega^2+k^2  \tilde c^2+\tilde c^2 q=-\tilde c^2 \sigma(k), $$
and we have that:
$$\omega=\pm c \sqrt{\tilde k^2 +q+\tilde \sigma(k)}. $$
Now, of course $\tilde \sigma(k)$ depends on the distribution of non-local links that we inserted in the wavefunction of the graph.\\\ \\ 
\textit{Two specific distributions.} Let us consider two specific cases:
\begin{itemize}
 \item $\rho_1(x-y)=\pi^{\frac{1}{4}}\ \sqrt{l}\ e^{-\frac{(x-y)^2}{2l^2}}$;
 \item $\rho_2(x-y)=\sqrt{2\ l}\ e^{-\frac{|x-y|}{2 l} }$.
\end{itemize}
In these cases we find, using standard tables of Fourier transforms:
\begin{itemize}
 \item $\tilde \sigma_1(k)=e^{-\frac{k^2}{a}}$;
 \item $\tilde \sigma_2(k)=\frac{a}{a^2 + k^2}$.
\end{itemize}
and thus, keeping track of all the factors, we obtain:
\begin{equation}
\omega_1=\pm \frac{1}{c} \sqrt{\frac{k^2}{1+S^2(K,l\ L)}+\tilde F_1(K,l\ L)+O(K,l\ L)e^{-\frac{k^2 l^2}{8}}} ,
\end{equation}
and
\begin{equation}
\omega_2=\pm \frac{1}{c}\sqrt{\frac{k^2}{1+S^2(K,l\ L)}+\tilde F_2(K,l\ L)+\frac{2 O(K,l\ L)}{\pi} \frac{ l^2}{l^2 + k^2}}, 
\end{equation}
with $\tilde F_1(K,l\ L) =\sqrt{2}\ O^2(K,l\ L)$ and $\tilde F_2(K,l\ L) = O^2(K,l\ L)$, which can be calculated by evaluating $\int_L \rho_i^4(x-y) dx$.
We have that $S^2(K,l\ L)=4 \tilde K^4 \xi^2/L^4$ and thus can be neglected with respect to $1$. Also, since the $c$ contribute with a factor of $L^2$ within the square root,
also $S^2$ can be neglected, and it contributes only the mass term in the  $L$.
Now we note  a nice property: both the two distributions go to $0$ for $k\rightarrow\infty$, that is, at high energy the dispersion relations
become Lorentz again. We see then that the total effect the one of having an effective scale-dependent mass, which runs from one mass to another one, in both cases:
\begin{equation}
 m_1(k)= \frac{1}{l\ L}\sqrt{S^2(K,l\ L)+S(K,l\ L)e^{-\frac{k^2 l^2}{8}}},
\end{equation}
\begin{equation}
 m_2(k)= \frac{1}{l\ L}\sqrt{S^2(K,l\ L)+S(K,l\ L) \frac{2 }{\pi} \frac{ l^2}{l^2 + k^2}}.
\end{equation}
The masses which are intertwined are given by
\begin{equation}
 m_1(0)= \frac{1}{l\ L}\sqrt{S^2(K,l\ L)+S(K,l\ L)},\ \ \ \  m_1(\infty)= \frac{S(K,l\ L)}{l\ L}, 
\label{m1}
\end{equation}
\begin{equation}
 m_2(0)= \frac{1}{l\ L}\sqrt{S^2(K,l\ L)+\frac{2 }{\pi}S(K,l\ L)},\ \ \ \  m_2(\infty)= \frac{S(K,l\ L)}{l\ L}. 
\label{m2}
\end{equation}
This property, of intertwining two different masses between $k=0$ and $k=\infty$ is shared by any function which is at least $\mathscr C^1$. It is remarkable, instead, 
that the mass at $k=\infty$ does not depend on the distribution we inserted at hand. In fact, any $\mathscr C^r$ distribution will lead to a Fourier
transformed distribution which goes to zero at $k=\infty$ as $1/k^r$ and thus tend to a finite value for the mass. 
Note that, if we send $l\rightarrow 0$, as required to have $S$ finite,
the dependence on the scale seems to disappear, leaving a Lorentz dispersion relation with a mass which depends on the function $S$. However, we have to remember that, in fact
these Fourier distributions come from the discrete dispersion relation. There, the distributions depend on $2\pi k/L$. If we define $\tilde k L= k$, then we have that the distributions
cancel out the dependence on $L$, leaving exactly (\ref{m1}) and (\ref{m2}) but dependent on this new momentum $\tilde k$.
Still, this mass depends on the distribution we have chosen through 
$\tilde \sigma_i(\tilde k=0)$ and so it has a valuable effect. 
We plot the running of $(l\ L) m_i (\tilde k)$ as a function of $x=K^2\ l\ L$, for the case $L=1$ in Fig. (\ref{running}) and $L=\infty$ in Fig. (\ref{running2}).

\begin{figure}
\centering
 \includegraphics[scale=0.5]{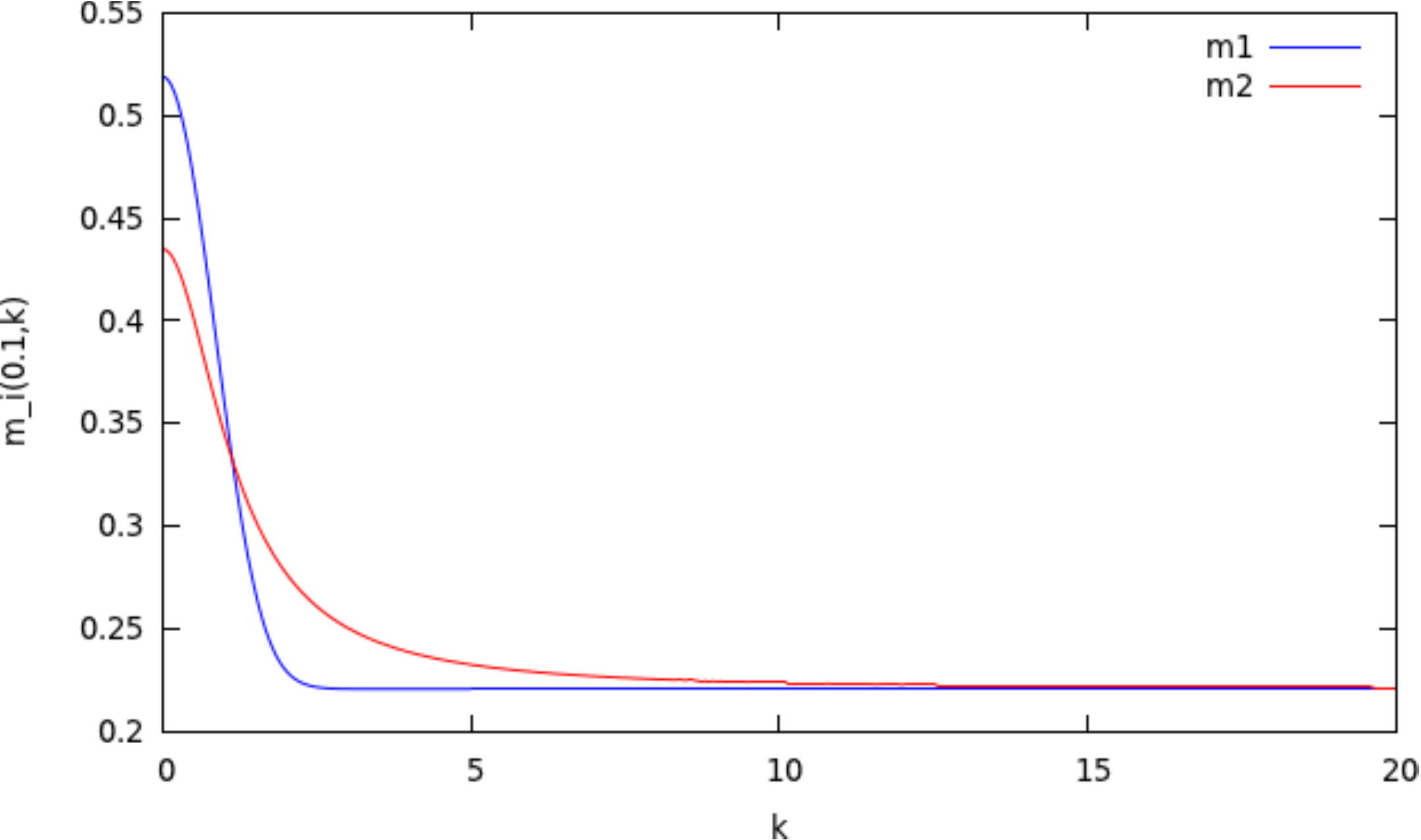}
\caption{The running of $m_i(\tilde k)$ for $x=0.1$, $l=1$ and $L=1$.}
\label{running}
\end{figure}
\begin{figure}
\centering
 \includegraphics[scale=0.5]{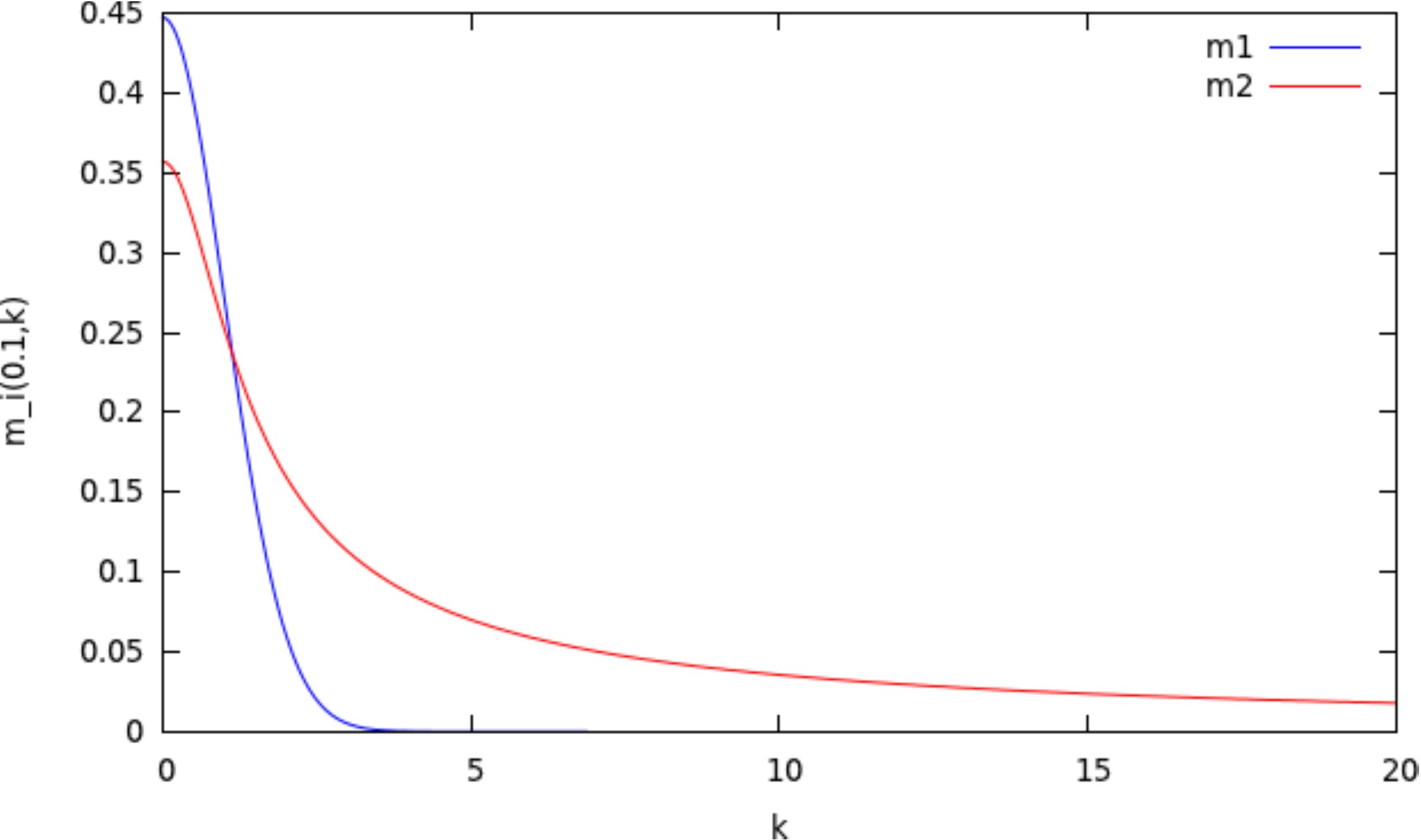}
\caption{The running of $m_i(\tilde k)$ for $x=0.1$, $l=1$ and $L=\infty$.}
\label{running2}
\end{figure}

We would like to point out that the appearence of a mass which is square-positive is rather surprising. The physical reason is that, before starting the calculation, we would 
have expected
that the presence of these non-local links would have shown a superluminal effect due to the non-local links themselves. However, the effective speed of propagation is higher 
because of the \textit{superposition} of
the graphs and not the non-local links. Indeed, the non-local links contributed \textit{only} in the mass, thus the term $C_z(t)$ additional to the differential equation we 
obtained. Besides, this mass is square-positive, thus it is
an effective mass and not a tachyonic one, which we would have expected from the presence of non-local links on physical grounds. The fact that it is square positive comes merely 
from the fact that the equation comes from a quantum mechanical average,
and thus the terms appear squared.
\section{Conclusions} 
One of the most striking theoretical consequences of General Relativity is the existence of wormholes and black holes. 
While the second is currently investigated experimentally,
less is known about the first. Here we discussed something which in principle is very similar, quantum states which violate macro-locality.
Besides, it could be that the quantum state of the Universe is
a superposition of spacetimes with non-local links. In the present paper we considered such a possibility in a toy model constructed using the framework Quantum Graphity.
In order to do so, we had to extend the results of \cite{graphity3} to a case in which the quantum state of the background is a superposition of many graph states. 
The superposition of these graphs was chosen so that it is 
dominated by a graph on which, as 
we showed in earlier papers, the expectation values of number operators of the bosons hopping on it satisfy a closed equation for probability density in the classical 
regime, i.e. a wave equation.
We extended the formula previously obtained and studied a particular case: graphs which violate micro- and macro- locality. As discussed, a violation of 
macro-locality can be interpreted, within the model, as the presence of spatial non-local links in the background spacetime. This 
is a concrete example of a quantum foam
within the framework of Quantum Graphity \cite{graphity1,graphity2}.
The graph state was chosen on the basis of what we know from low energy physics, which is that Lorentz invariance is satisfied up and above the Planck scale\cite{Planckdata}. 
We also used a class of graphs introduced in \cite{graphity3}, rotationally invariant graphs. By exploiting their symmetry, the problem can be reduced to a 1-dimensional
one, i.e. Bose-Hubbard model on a line with specific couplings depending on the connectivity of the graph.
We thus constructed the states that are corrections to the low-energy physics by assuming that the non-local links are suppressed by a length 
according to a certain
distribution. The length is measured by a combinatorial distance based on the \textit{low energy} graph and which defines the state. We studied 
for the cases $d(x,y)=(x-y)^2$
and $d(x,y)=|x-y|$. We found that, in the continuum limit, there is no superluminal effect on the low-energy physics, i.e. the speed of propagation is intact. However, 
there is an appearance of a mass dependence
on the constants of the distribution and that can be calculated within the model. These masses are square-positive and thus do not violate the physics 
of the restricted Lorentz group, i.e., 
are not tachyonic. A simple analysis showed that this mass runs with the energy scale and, in particular, runs to zero at high energy. It is interesting to ask whether a 
similar phenomenon happens
for the other fields. This analysis suggests the possibility that a quantum foam could contribute to the mass of a quantum field. As suggested in \cite{marksm} and \cite{smochan}, 
the possibility of having non-local link states within Loop Quantum Gravity is very natural. Also, it has been suggested that these states could contribute to the dark energy puzzle.
The results of the present paper suggests that, as in \cite{smochan}, the quantum foam contributes to the mass of fields hopping on such a superposition of spacetimes. We believe
that such possibility needs to be further investigated.
 
\newpage
\begin{center}
{\bf Aknowledgements}\\ 
\end{center}
Research at Perimeter Institute is supported by the Government of Canada
through Industry Canada and by the Province of Ontario
through the Ministry of Research \& Innovation. This research has been made possible by financial support of the Templeton and Humboldt Foundations.


\begin{thebibliography}{99}
\bibitem{WheelerFord} J.A. Wheeler, K. Ford, Geons, black holes and quantum foam: a life in physics, W.W. Norton Company, Inc., New York (1998) 
\bibitem{lw} M. Visser, Lorentzian Wormholes: from Einstein to Hawking, American Institute of Physics Press (Woodbury, New York) 1992. 
\bibitem{graphity1}
  T. Konopka, F.Markopoulou and L. Smolin,
  arXiv:hep-th/0611197 ;
  T. Konopka, F. Markopoulou, S. Severini, Phys. Rev. D 77, 104029 (2008),
arXiv:0801.0861;
  F. Caravelli, F. Markopoulou, Phys. Rev. D 84 024002 (2011), arXiv:1008.1340
\bibitem{graphity2}
A. Hamma, F. Markopoulou, S. Lloyd, F. Caravelli, S. Severini, K. Markstrom, Phys. Rev. D 81, 104032 (2010), 
arXiv:0911.5075
\bibitem{graphity3} F. Caravelli, A. Hamma, F. Markopoulou, A. Riera,  arXiv:1108.2013
\bibitem{konopka} T. Konopka, Phys. Rev. D78 044032 (2008), [arXiv:0805.2283 [hep-th]].
\bibitem{stleomaz} K.S. Stelle, Phys.Rev. D16 (1977) 953-969; L. Modesto, arXiv:1107.2403; T. Biswas, E. Gerwick, T. Koivist, A. Mazumdar, arXiv:1110.5249;
\bibitem{marksm} F. Markopoulou, L. Smolin, Class.Quant.Grav. 24 (2007) 3813-3824, arXiv:gr-qc/0702044.
\bibitem{loop} C. Rovelli, Quantum Gravity, Cambridge University
Press, Cambridge (2004); A. Ashtekar, Class. Quant.
Grav. 21, R53 (2004), arXiv:gr-qc/0404018; T. Thiemann, [gr -qc/0110034]; A. Perez, arXiv:gr-qc/0409061; 
\bibitem{causets} R. Sorkin, Proceedings of the
Valdivia Summer School, edited by A. Gomberoff, D. Marolf, arXiv:gr-qc/0309009.
\bibitem{revqg} F. Markopoulou, A. Hamma,  New J. Phys. 13:095006 (2011), arXiv:1011.5754;
\bibitem{smochan} C. Prescod-Weinstein, L. Smolin, Phys. Rev. D80 063505 (2009), arXiv:0903.5303.
\bibitem{florian} F. Conrady, J.Statist.Phys.142:898 (2011), arXiv:1009.3195
\bibitem{Planckdata} Planck collaboration, Nature Physics 462, 331-334 (2009);
%
%




\end{thebibliography}
\end{document}